\documentstyle[aps,prl,floats]{revtex}
\begin{document}
\draft

\twocolumn[\hsize\textwidth\columnwidth\hsize\csname @twocolumnfalse\endcsname

\title{Macroscopic parity violating effects and
$^3$He-A }
\author{ G.E. Volovik$^{1,2}$ and A. Vilenkin$^{3}$}
\address{
$^{1}$Low Temperature Laboratory, Helsinki University of
Technology,
P.O.Box 2200, FIN-02015 HUT, Finland\\
$^{2}$L.D. Landau Institute for
Theoretical Physics,, Kosygin Str. 2, 117940 Moscow, Russia\\
$^{3}$Institute of Cosmology,
Department of Physics and Astronomy,
Tufts University,
Medford, MA 02155
USA}

\date{\today}
\maketitle

\begin{abstract}
We discuss parity violating effects in relativistic quantum
theory and their analogues in effective field theory of
superfluid $^3$He-A.  A mixed axial-gravitational Chern-Simons
term in the relativistic
effective action and its condensed matter analog are
responsible for
the chiral fermion
flux along the rotation axis of the heat bath in relativistic
system and for the unusual $ {\bf \Omega}$-odd dependence of the
zero-temperature
density of the normal component on the rotation velocity in $^3$He-A.
\end{abstract}

\

\pacs{PACS numbers: 11.30.Er, 11.15.-q, 67.57.-z, 98.80.Cq}

\

] \narrowtext

\section{Introduction}

In modern view, relativistic quantum field theory and general
relativity may be emergent phenomena arising in the low energy corner of the
quantum vacuum as effective theories of vacuum fluctuations
\cite{Weiberg,Sakharov}. Such effective theories of the collective degrees of
freedom are typical for condensed matter systems. In a particular
universality class of
systems, the low energy
properties are very similar to those of the relativistic quantum vacuum.
Superfluid $^3$He-A is a representative of this class
\cite{PNAS}.   Lacking practically  any symmetries above the superfluid
transition, $^3$He-A in the extreme limit of low energy acquires most of the
symmetries known today in particle physics: (analogs of) Lorentz
invariance, gauge invariance, general covariance, etc. The analogs of chiral
Weyl fermions, as well as of gauge bosons and gravity field appear as fermionic
and bosonic collective modes of the $^3$He-A ground state.  Such conceptual
similarity between the condensed matter of this class and the quantum vacuum
makes $^3$He-A an ideal laboratory for simulating relativistic field theory
effects in high energy physics and cosmology.

Parity violation is one of the fundamental properties of the quantum
vacuum. This effect is strong at high energy of the order of electroweak
scale, but is almost imperceptible in the low-energy physics. For
example, Leggett's suggestion to observe the macroscopic
effect of parity violation using such macroscopically coherent atomic system
as superfluid $^3$He-B is very far from realization
\cite{Leggett,VollhardtWolfle}. On the other hand,   an analog of parity
violation exists in superfluid $^3$He-A alongside with the related phenomena,
such as chiral anomaly and macroscopic chiral currents (for a review see Refs.
\cite{VollhardtWolfle,Exotic}). So, if we cannot investigate the macroscopic
parity violating effects directly we can simulate analogous
physics in $^3$He-A.

The reason why all the attributes of relativistic quantum field theory
arise in $^3$He-A
can be traced to the existence of gap nodes, stable point zeros
in the fermion energy spectrum.
Close to each gap node the fermions necessarily become Weyl fermions and can be
described by a
general $2\times 2$ matrix Hamiltonian
\begin{equation}
{\cal H}= e^i_a (p_i - p_{0i})\tau^a ~.
\label{Weyl}
\end{equation}
Here the three vectors $e_i^a$ play the role of a dreibein field which gives
rise to the effective gravity field, while the position of the node in the
momentum space, $p_{0i}$,  plays the role of the effective
electromagnetic field. These fields  are dynamical and
the low energy dynamics of these collective modes is governed by the effective
action, obtained by integration over the fermionic degrees of freedom, which is
in complete analogy with the effective gravitational and
electromagnetic actions
introduced by Sakharov \cite{Sakharov} and Zeldovich \cite{Zeldovich}
respectively. If the main contribution to the integral comes from the low
energy
``relativistic'' fermions, the effective action automatically adopts the gauge
invariance and general covariance of the low energy fermionic Lagrangian.

The energy spectrum of fermions in $^3$He-A is
\begin{eqnarray}
\hat H_A = {\tau}^3 {p^2 -p_F^2\over 2m^*} + {\Delta_0^2\over p_F^2}[{\tau}^1
{\bf p}\cdot \hat{\bf e}_1 + {\tau}^2 {\bf p}\cdot \hat{\bf e}_2]  ~,
\label{Nambu}\\
 E^2({\bf p})  = \hat H_A^2=
 \left({p^2\over 2m^*}-{p_F^2\over 2m^*}\right)^2
             +{\Delta_0^2\over p_F^2}({\bf p}\times \hat{\bf l})^2\,.
\label{Esquared}
\end{eqnarray}
Here $\Delta_0$ is the amplitude of the gap; $m^*$  is the effective mass
of the fermionic quasiparticles in the normal Fermi liquid, where
$\Delta_0=0$ and the spectrum is quadratic: $E({\bf p})  =
{(p^2-p_F^2)/2m^*}$;
$p_F$ is
the Fermi momentum of the normal Fermi liquid; $\hat{\bf l}$ is a unit
vector which
determines the direction of the orbital angular momentum of the Cooper pairs;
$\hat{\bf e}_1$ and
$\hat{\bf e}_2$ are unit mutually orthogonal vectors with $\hat{\bf e}_1 \times
\hat{\bf e}_2 =\hat{\bf l}$. This spectrum contains two gap nodes,  at
${\bf p}_0= \pm p_F\hat{\bf l}$, and close to each of these nodes the
``relativistic'' equation (\ref{Weyl}) is approached.

The transverse modes related
with the dynamics of the
$\hat{\bf l}$ vectors are Goldstone bosons known as orbital waves. Since in the
relativistic domain the momentum shift
${\bf  A}^{\rm eff}=p_F\hat{\bf l}$ plays the role of the effective
electromagnetic field acting on quasiparticles, the orbital waves represent
photons  and Eq.(\ref{Weyl}) can be rewritten as
\begin{equation}
{\cal H}= e^i_a (p_i - eA^{eff}_i)\tau^a ~
\label{Weyl'}
\end{equation}
with $e=\pm 1$.

The fermonic quasiparticles living in the vicinity of
opposite nodes have opposite chirality:  The left-handed particles have
positive
``charge'', $e_L=+1$, with respect to the effective field ${\bf  A}^{\rm
eff}$, i.e. they are concentrated near ${\bf p}_0= + p_F\hat{\bf l}$. The
negatively charged quasiparticles, i.e. those near ${\bf p}_0= -
p_F\hat{\bf l}$,  are right-handed  fermions, $e_R=-1$ \cite{Exotic}. The
Pauli matrices in $^3$He-A, $\tau^a$, which play the role of spin of the
fermion and thus determine its chirality, are actually defined in the
Bogoliubov-Nambu particle-hole space and thus describe the Bogoliubov spin. On
the contrary, the ordinary spin of the $^3$He atoms, which is not introduced in
Eq.(\ref{Nambu}) being irrelevant for our consideration, corresponds to the
weak isospin and gives rise to the effective $SU(2)$ gauge field.

\section{Axial anomaly in $^3$He-A.}

Massless chiral fermions give rise to a number of anomalies in the
effective action. The advantage of $^3$He-A is that this system is
complete: not
only the ``relativistic'' infrared regime is known, but also the behavior
in the
ultraviolet ``nonrelativistic'' (or ``transplanckian'') range is calculable, at
least in principle. Since there is no need for a cut-off, all subtle issues of
the anomaly can be resolved on physical grounds. The measured quantities
related
to the anomalies depend on the correct order of imposing limits, i.e. on what
parameters of the system tend to zero faster: temperature $T$;
external frequency $\omega$; inverse
quasiparticle lifetime due
to collisions with thermal fermions $1/\tau$; inverse volume; the distance
$\omega_0$ between the energy levels of fermions, etc.  All this is very
important
for the $T\rightarrow 0$ limit, where   $\tau$ is formally infinite.
An example of
the crucial difference between the results obtained  using
different limiting procedures is
the so called  ``angular momentum paradox''  in
$^3$He-A, which is also related to the anomaly: The orbital
momentum of the fluid at $T=0$ differs by several orders of magnitude,
depending on
whether the limit is taken while keeping $\omega\tau\rightarrow 0$ or
$\omega\tau\rightarrow \infty$. The ``angular momentum paradox'' in
$^3$He-A has
possibly a common origin with the anomaly in the spin structure of hadrons
\cite{Troshin}.

In the spatially inhomogeneous
case there is another important parameter
$\omega_0\tau$,  where $\omega_0$ is the distance between the localized
energy levels. The gapless fermions in $^3$He-A lead to the momentum exchange
between the superfluid vacuum and the normal component of the liquid (the gas
of fermionic quasiparticles). This exchange is mediated by the texture of the
$\hat{\bf l}$ field:
${\dot{\bf P}}= (p_F^3/2\pi^2)\hat{\bf l}
\left(\partial_t\hat{\bf l}\cdot(\nabla\times \hat{\bf l})\right)$. Since the
dictionary for translation to the language of relativistic theory reads
${\bf  A}^{\rm eff}\equiv p_F\hat{\bf l}$, one obtains
\begin{eqnarray}
{\dot{\bf P}}=p_F\hat{\bf
l}~\dot n~,
\label{MomentumProduction}\\
\dot n= ~{1\over 2\pi^2}
 \partial_t {\bf A}^{\rm eff}\cdot(\nabla\times  {\bf A}^{\rm eff})~.
\label{AdlerBellJackiw}
\end{eqnarray}
Eq.(\ref{AdlerBellJackiw}) is nothing but the Adler-Bell-Jackiw axial
anomaly
equation  \cite{Adler}, describing the production
of chiral fermions from the quantum vacuum due to the spectral flow through the
gap nodes.  The relevant fermionic charge, which is produced due to the chiral
anomaly in $^3$He-A, is the linear momentum $p_F\hat{\bf l}$ of fermionic
quasiparticle: It is the rate of the momentum production, which is measured in
experiments on the dynamics of the vortex textures as an extra force acting
on a
moving texture
\cite{Bevan}.

It appears, however, that
Eqs.(\ref{MomentumProduction}-\ref{AdlerBellJackiw}) are valid only in the
limit of
continuous spectrum, i.e. when the distance
$\omega_0$ between the energy levels of fermions in the texture is much
smaller   than the inverse quasiparticle lifetime: $\omega_0\tau \ll 1$.  The
spectral flow completely
disappears in the opposite case $\omega_0\tau \gg 1$, because the spectrum
becomes effectively discrete.  As a result, the force acting on a vortex
texture differs by
several orders of magnitude for the cases
$\omega_0\tau \ll 1$ and $\omega_0\tau \gg 1$. The parameter $\omega_0\tau$ is
regulated by temperature.  The Adler-Bell-Jackiw equation was experimentally
confirmed in experiments with rotating
$^3$He-A performed in the limit $\omega_0\tau \ll 1$
\cite{Bevan,LammiTalk}.
The transfer from the axial anomaly regime $\omega_0\tau \ll
1$ to the regime of the suppressed spectral flow $\omega_0\tau \gg 1$ has been
observed for $^3$He-B vortices, whose dynamics is governed by
the similar spectral flow in the vortex core \cite{Bevan,LammiTalk}.

The chiral anomaly also leads to the chiral current, which is proportional to
${\bf  A}^{\rm eff}\cdot(\nabla\times {\bf  A}^{\rm eff})$. This current
has been
proved to exist: it leads to an observed instability of the superflow
\cite{Experiment,LammiTalk}. The same instability of the system
of right-handed electrons towards production of a hypermagnetic field was
discussed by Joyce and Shaposhnikov \cite{JoyceShaposhnikov} in relation
to the generation  of a primordial magnetic field.

\section{ Mixed axial-gravitational Chern-Simons term in the effective
action.}

\subsection{Parity violating current}

Here we discuss another particular case of correspondence between
relativistic field theory, in which the chiral anomaly
problem can be mapped to the angular momentum paradox in
$^3$He-A. It involves macroscopic parity violating effects in a rotating system
with chiral fermions, discussed in \cite{Vilenkin79}.
The angular velocity of rotation ${\bf \Omega}$
defines the preferred direction of
polarization, and right-handed fermions move in the direction
of their spin.  As a result, such fermions develop a current
parallel to ${\bf \Omega}$.  Similarly, left-handed fermions
develop a current antiparallel to ${\bf \Omega}$.  The corresponding
current density was calculated in \cite{Vilenkin79}, assuming thermal
equilibrium at temperature $T$ and chemical potential of the fermions
$\mu$.  For right-handed fermions, it is given by
\begin{equation}
{\bf j}=\left({T^2\over{12}}+{\mu^2\over{4\pi^2}}\right){\bf\Omega}.
\label{PVC}
\end{equation}
The current ${\bf j}$ is a polar vector, while the angular velocity
${\bf\Omega}$ is an axial vector, and thus Eq.(\ref{PVC}) violates the
reflectional symmetry.  We are going to show that this current gives
rise to what can be called mixed axial-gravitational Chern-Simons
terms in the effective action and that equivalent terms do exist in
the thermodynamic potential of $^3$He-A.

If the current (\ref{PVC}) is coupled to a
gauge field $A^\nu$, the appropriate term in the Lagrangian density is
\begin{equation}
 L =e{\bf A\cdot j}/c^2,
\label{calL}
\end{equation}
where $e$ is the gauge coupling.
The correspondence between field theory and $^3$He-A is
achieved by replacing the gauge field and the metric
by appropriate $^3$He-A observables \cite{Exotic}.  Here we shall
start from the opposite end and derive the ${\bf \Omega}$-dependent
contribution to the free energy.  We shall then show that it is
equivalent to (\ref{calL}).  We note that to establish the
correspondence, the free energy should be expressed in a covariant and
gauge invariant form and should not contain any
material parameters, such as ``speed of light''.  Then it can be equally
applied to both systems, standard model and $^3$He-A.

\subsection{Orbital angular momentum and free energy}

Let us consider a stationary liquid $^3$He-A in a vessel rotating with
angular velocity ${\bf\Omega}$ at a nonzero temperature.  We assume a
spatially homogeneous vector ${\hat{\bf l}}$ oriented along the
rotation axis. In $^3$He-A this can be achieved in the parallel-plane
geometry, while in the layered oxide
superconductor Sr$_2$RuO$_4$, which is believed to be a triplet superconductor
with a $^3$He-A-like order parameter,   the ${\hat{\bf
l}}$-vector is always fixed along the normal to the layers  \cite{Rice}.
The superfluid component (vacuum) is assumed to be at rest,  while the
normal component -- the heat bath of thermal fermions -- circulates in
the plane perpendicular to
$\hat{\bf l}$ with the velocity ${\bf v}_{\rm n}={\bf\Omega} \times {\bf r}$.

The value of the angular momentum of a rotating $^3$He-A has been a
subject of a long-standing controversy (for a review see
\cite{VollhardtWolfle,Exotic}).
Different methods for calculating the angular momentum give results
that differ by many orders of magnitude.  The result is also sensitive
to the boundary conditions, since the angular momentum in the liquid is not
necessarily the local quantity, and to whether the state is strictly
stationary or
has a small but finite frequency.  This is often referred to as the angular
momentum paradox.  The paradox is related to the axial anomaly induced by
chiral
quasiparticles and is now reasonably well understood.

According to Kita conjecture \cite{Kita}, which was supported by
his numerical calculations, the total angular momentum of the liquid with
${\hat{\bf l}}=const$ corresponds to the following angular momentum
density
\begin{equation}
 {\bf L}(T)={\hbar\over 2} ~\hat{\bf l}~ n_{{\rm s}\parallel}(T)~,
 \label{MomentumDensity}
\end{equation}
where  $n_{{\rm s}\parallel}(T)$ is the temperature dependent density of the
superfluid component when it flows along   $\hat{\bf l}$. We recall
that the  current of the $^3$He
atoms has two contributions in the superfluid state:
\begin{equation}
{\bf J}= n {\bf v}_{\rm s}+
{\bf J}_{\rm q}~,~{\bf J}_{\rm q}=\sum_{\bf p} {{\bf p} \over m_3} f({\bf p})~.
\label{TotalCurrent}
\end{equation}
The first term is the current transferred by the superfluid
vacuum moving with velocity ${\bf v}_{\rm s}$; $n$ is the particle density  of
$^3$He liquid. The 2nd term is the contribution of  quasiparticles,  where
$f({\bf p})$ is the quasiparticle distribution function and  $m_3$   is the
bare
mass of $^3$He atom.  In equilibrium one has
\begin{equation}
f({\bf p})=\left(\exp { \tilde E({\bf p})+ {\bf p}  {\bf v}_n\over
T} +1\right)^{-1},~\tilde E=E({\bf p})+ {\bf
p}\cdot{\bf v}_s~,
\label{Equilibrium}
\end{equation}
where $\tilde E$ is the Doppler shifted energy of quasiparticle, when the
superfluid vacuum is moving. In the linear in velocity regime one obtains
$ J_{{\rm q} i} =n_{{\rm n}ik}(v_{{\rm n}k}-v_{{\rm s}k})$, where
$n_{{\rm n}ik}$ is the so called density of the normal component. In
$^3$He-A the normal component density is a
uniaxial tensor with the anisotropy axis along
$\hat{\bf l}$: the
density involved in the normal motion depends on the orientation
of ${\bf v}_{\rm n}-{\bf v}_{\rm s}$ with respect to
$\hat{\bf l}$ \cite{VollhardtWolfle}. The tensor of the superfluid
density is $n_{{\rm s}ik}=n\delta_{ik}-n_{{\rm n}ik}$. At
$T=0$ one has
$n_{{\rm s}\parallel}(0)=n$.

The contribution of Eq.(\ref{MomentumDensity}) to the free energy
density is
\begin{equation}
   F= -{\bf\Omega} \cdot {\bf L}(T)=-{\bf\Omega} \cdot {\bf L}(0)+
{\hbar\over 2} ({\bf\Omega}\cdot \hat{\bf l})  n_{{\rm n}\parallel}(T)~.
 \label{EnergyDensity}
\end{equation}
  The first (zero-temperature) term on the right-hand side of
(\ref{EnergyDensity}) has no analogue in field theory, and we disregard it
in what follows. The 2nd term  comes from chiral
quasiparticles, which comprise the normal component. Its longitudinal
density $n_{{\rm n}\parallel}=n - n_{{\rm s}\parallel}$ at $T\ll T_c$
is obtained from Eqs.(\ref{Equilibrium}-\ref{TotalCurrent})
\cite{VollhardtWolfle}:
\begin{equation}
 n_{{\rm n}\parallel}
\approx { m^*\over 3m_3} p_F^3
{T^2\over \Delta_0^2} ~~.
\label{LowTnormalDensity}
\end{equation}
 Kita \cite{Kita}  obtained his
result Eq.(\ref{MomentumDensity}) for the
simplest case when $m^*=m_3$, i.e. the interactions which
renormalize the quasiparticle mass    in normal Fermi  liquid were
neglected. We expect, however, that Eq.(\ref{MomentumDensity}) is
insensitive to the details of interaction (see below) and shall
concentrate on this case.

\subsection{Effective
Chern-Simons action:  $T\neq 0$, $\mu=0$}

Let us now use the dictionary for translating Eq.(\ref{EnergyDensity})
to the language of relativistic theories.  If one chooses the reference frame
rotating with the normal component, then the effective gauge field and
the effective
metric ``seen'' by the fermionic quasiparticles are
\begin{eqnarray}
{\bf A}\equiv{\bf  A}^{\rm eff}= p_F\hat{\bf l}~,\\
 g^{ik}=c_\perp^2 (\hat e_1^i \hat e_1^k
+  \hat e_2^i \hat e_2^k)+c_\parallel^2 \hat l^i \hat l^k -v_{\rm n}^i v_{\rm
n}^k~,~g^{0i}=v_{\rm n}^i\\  c_\perp={\Delta_0\over
p_F}~,~c_\parallel={p_F\over m_3}
\label{RelSp}
\end{eqnarray}
They are obtained by linearizing Eq.(\ref{Nambu}) in the vicinity of the
nodes. The mixed components of the metric tensor $g^{0i}=v_{\rm n}^i$
come from
the Doppler shift in the comoving frame.  The angular
velocity is expressed through
${\bf v}_{\rm n}$ as
${\bf\Omega} =(1/2){\bf\nabla}\times {\bf v}_{\rm n}$ and thus
is proportional to the effective gravimagnetic field
\cite{GravimagneticMonopole}
\begin{equation}
{\bf B}_g={\bf\nabla}\times {\bf g}=2{{\bf\Omega}\over c_\perp^2}~,  ~{\bf
g}\equiv
g_{0i}= {v_{{\rm n}i}\over c_\perp^2}.
 \label{GravimagneticField}
\end{equation}
Here, we have made the following assumptions: (i) $v_{\rm n}\ll
c_\perp$ everywhere in the vessel, i.e. the counterflow velocity ${\bf
v}_{\rm n} -{\bf v}_{\rm s}$ is smaller
than the pair-breaking critical velocity $c_\perp=\Delta_0/p_F$ (the transverse
"speed of light"). This means that
there is no  region in the vessel where
particles can have negative energy (ergoregion). Effects caused by the
ergoregion in superfluids are discussed in
\cite{CalogeracosVolovik}.
(ii)  Rotation velocity is so small that there are no vortices in the
container. This is typical for superfluid $^3$He, where the critical velocity
for nucleation of vortices is comparable to the pair-breaking velocity.
Even in the geometry when the $\hat{\bf l}$-vector is not fixed, the
observed critical velocity in $^3$He-A was found to reach 0.5
rad/sec. For the geometry with fixed $\hat{\bf l}$, it should be
comparable with the critical velocity in $^3$He-B.   (iii)  We approach the
$T\rightarrow 0$ limit in such a way that there is still a reference frame
specified by the thermal bath of fermionic excitations, which rotates
together with the container in equilibrium. This corresponds to the case when
the condition $\omega\tau \ll 1$ remains valid, despite the divergence
of $\tau$.

The translation of Eqs.(\ref{EnergyDensity}-\ref{LowTnormalDensity}) to
the relativistic language
can be presented as the following term in the Lagrangian:
\begin{equation}
 L_{\rm CS}= {e_R-e_L \over 24} ~ T^2{\bf A}\cdot {\bf B}_g= {e_R-e_L \over 24}
~T^2 e^{ijk} A_i
\nabla_j g_{k0}~.
 \label{ChernSimons1}
\end{equation}
 Since the
parity of $^3$He-A fermions is opposite to their charge, the contributions
of the two species in $^3$He-A are not cancelled but are added together.

Eq.(\ref{ChernSimons1}) does not contain explicitly any material parameters,
such as the "speeds of light" $c_\parallel$ and $c_\perp$.  Moreover,
it is gauge invariant,
provided that the system is in thermal equilibrium, i.e. $T=const$.  Thus
it can
be immediately applied to the relativistic theory in a rotating frame
in Minkowski space,
where $e^{ijk} \nabla_j
g_{k0}=2\Omega^i/c^2$, with $c$ being the speed of light. Note that the angular
velocity ${\bf\Omega}$ retains the same meaning in relativistic theory.
It is the
angular velocity which appears in the distribution function of thermal
fermions.
With the aid of Eq.~(\ref{PVC}) it is easily verified that
Eq.~(\ref{ChernSimons1}) is equivalent to the Lagrangian density
(\ref{calL}), if we set $\mu=0,~e_R=e$ and $e_L=0$.

Eq.(\ref{ChernSimons1}) is not Lorentz invariant, but this is not
important here
because the existence of a heat bath
does violate the Lorentz invariance, since it provides a distinguished
reference frame. To restore the Lorentz invariance and also the general
covariance one must introduce the 4-velocity ($u^{\mu}$) and/or  4-temperature
($\beta^{\mu}$) of the heat bath fermions. But this is not necessary
since the unification of the chiral effects in the two systems has been
achieved already at this level.

We next consider the effect of a finite chemical potential,
which in our case
corresponds to the superfluid-normal counterflow.  We shall assume a
superflow along the axis of the cylinder and consider
its effect in the presence of rotation.

\subsection{Nonzero $\mu$ vs nonzero axial counterflow.}

A superfluid-normal counterflow velocity along ${\hat{\bf l}}$ produces
a Doppler shift, $\tilde E = E + {\bf p}\cdot({\bf v}_{\rm s}-{\bf v}_{\rm
n})$. In the vicinity of the two nodes one has $\tilde E\approx E\pm
p_F\hat{\bf
l}\cdot({\bf v}_{\rm s}-{\bf v}_{\rm n})$, which means that the counterflow
enters the Lagrangian for fermionic quasiparticles in
$^3$He-A in the same way as the chemical potentials for relativistic chiral
fermions \cite{LammiTalk}:
\begin{equation}
\mu_R= -\mu_L\equiv - p_F{\hat{\bf l}}\cdot({\bf v}_{\rm s}-{\bf v}_{\rm n})
\label{ChemicalPotential}
\end{equation}
That is why the energy stored in the system of chiral fermions, $A=F-\mu_R N_R
-\mu_L N_L$, and the energy of the counterflow along the $\hat{\bf
l}$ vector,  $A=F - m_3({\bf J}_{\rm q}\cdot \hat{\bf
l})(\hat{\bf
l}\cdot ({\bf v}_{\rm n}-{\bf v}_{\rm s}))$, are described by the same
thermodynamic potential.  At
$T=0$ it is
\begin{equation}
 A= - { \sqrt{-g}\over 12\pi^2} (\mu_R^4+\mu_L^4)\equiv
-{m_3p_F^3\over 12 \pi^2 c_\perp^2} \left({\hat{\bf l}} \cdot({\bf v}_{\rm
s}-{\bf
v}_{\rm n})\right)^4  .
\label{CounterflowEnergy}
\end{equation}
Variation of
Eq.(\ref{CounterflowEnergy}) with respect to  ${\bf
v}_{\rm n}$ gives the mass current along the
${\hat{\bf l}}$-vector produced by the fermionic quasiparticles, which at
$T=0$ occupy the
energy levels with $\tilde E<0$:
\begin{equation}
 m_3  J_{{\rm q}\parallel}=  -{d  A \over
dv_{{\rm n}\parallel}}~~.
\label{ZeroTCountercurrent}
\end{equation}
This shows that in the presence of a superflow with respect to the heat bath
the normal component density is nonzero even  in the limit $T\to 0$
\cite{Muzikar1983}:
\begin{equation}
 n_{{\rm n}\parallel}(T\to 0)={d  J_{{\rm q}\parallel}\over dv_{{\rm
n}\parallel}} =
  {p_F^3\over 3\pi^2 c_\perp^2}  ({\hat{\bf l}} \cdot({\bf v}_{\rm s}-{\bf
v}_{\rm n}))^2
  ~~.
\label{ZeroTnormalDensity}
\end{equation}

Now we can check how general is the Kita conjecture,
Eq.(\ref{MomentumDensity}). Let us assume that
Eq.(\ref{EnergyDensity}) remains valid even if the normal component at
$T=0$ is added. Then from Eq.(\ref{ZeroTnormalDensity}) it follows
that the mixed Chern-Simons term becomes
\begin{equation}
 L_{\rm CS}={{e_R\mu^2_R-e_L\mu^2_L}\over {8\pi^2}}{\bf
A}\cdot {\bf B}_g~.
 \label{ChiralTerm2}
\end{equation}
This term is equivalent
to the
Lagrangian density (\ref{calL}),(\ref{PVC}) in the appropriate limit
($T=0,~e_R=e,~e_L=0, ~\mu_R=\mu,~\mu_L=0)$, which thus confirms the above
assumption.

\section{Discussion.}

The form of Eq.(\ref{ChiralTerm2}) is similar to that of the induced
Chern-Simons term
\begin{equation}
{\tilde L}_{CS}={\mu\over{2\pi^2}}{\bf A\cdot B}
\label{CS}
\end{equation}
with ${\bf B=\nabla\times A}$, which has been extensively discussed
both in the context of chiral fermions in relativistic theory
\cite{Vilenkin80,Redlich,JackiwKostelecky,Andrianov} and in $^3$He-A
\cite{LammiTalk}.  The main difference between (\ref{CS}) and
(\ref{ChiralTerm2}) is that ${\bf B_g=\nabla\times g}$ is the
gravimagnetic field, rather than the magnetic field ${\bf B}$
associated with the potential ${\bf A}$.  Hence the name ``mixed
Chern-Simons term''.

The parity-violating currents (\ref{PVC}) could be induced in
turbulent cosmic plasmas and could play a role in the origin of cosmic
magnetic fields \cite{Leahy}.  The corresponding liquid Helium effects
are less dramatic but may in principle be observable.

Although the mixed Chern-Simons terms have the same form in
relativistic theories and in $^3$He-A, their physical manifestations
are not identical.
In the relativistic case, the electric current of chiral
fermions is obtained by
variation with respect to
$ {\bf A}$, while in $^3$He-A case the observable effects are obtained by
variation of the same term but with respect to $^3$He-A observables.
For example, the
expression for the current of $^3$He atoms is obtained by variation of
Eq.(\ref {ChiralTerm2}) over ${\bf
v}_{\rm n}$. This leads to
an extra particle current along the rotation axis, which is odd in
${\bf\Omega}$:
\begin{equation}
\Delta {\bf J}_{\rm q}({\bf\Omega})= {p_F^3\over \pi^2 }  {\hat{\bf
l}}~({\hat{\bf l}}\cdot({\bf v}_{\rm s}-{\bf v}_{\rm n})) ~ {{\hat{\bf l}
}\cdot
{\bf\Omega}\over m_3 c_\perp^2}~.
 \label{Current}
\end{equation}
Eq.(\ref{Current}) shows that there is an ${\bf\Omega}$ odd contribution to
the normal component density at $T\to 0$ in $^3$He-A:
\begin{equation}
\Delta n_{{\rm n}\parallel}({\bf\Omega}) = {\Delta J_{\rm
q}({\bf\Omega})\over v_{{\rm n}\parallel}-v_{{\rm s}\parallel}}
=  {p_F^3\over \pi^2 }  {{\hat{\bf l}}\cdot
{\bf\Omega}\over  m_3 c_\perp^2} ~.
\label{NormalDensityInRotation}
\end{equation}
The sensitivity of the normal component density to the direction of
rotation
is the counterpart of the parity violation effects in relativistic
theories with chiral fermions.
It should be noted though that, since ${\bf{\hat l}}$ is an axial
vector, the right-hand sides of (\ref{Current}) and
(\ref{NormalDensityInRotation}) transform, respectively, as a polar
vector and a scalar, and thus (of course) there is no real parity violation
in $^3$He-A. However, a nonzero expectation value of the axial vector of  the
orbital angular momentum ${\bf L} =(\hbar/2)n_{{\rm s}\parallel}(T){\bf {\hat
l}}$  does indicate a {\it spontaneously} broken reflectional symmetry, and an
 internal observer ``living'' in a $^3$He-A  background with a fixed
 ${\bf{\hat l}}$ would observe parity-violating effects.

The contribution (\ref{NormalDensityInRotation}) to
the normal component density can have arbitrary sign depending on
the sense of rotation with respect to ${\hat{\bf l}}$. This however does not
violate the general rule that the overall normal component density must be
positive:  The rotation dependent current $\Delta{\bf J}_{\rm
q}(\vec\Omega)$ was calculated as
a correction to the rotation independent current in
Eq.(\ref{ZeroTCountercurrent}). This means that we used the condition
$\hbar\Omega \ll m_3 (v_{{\rm s}\parallel} - v_{{\rm n}\parallel})^2 \ll m_3
c_\perp^2$. Under
this condition the overall normal density, given by the sum of
(\ref{NormalDensityInRotation}) and
(\ref{ZeroTnormalDensity}), remains positive.

The ``parity'' effect in Eq.(\ref{NormalDensityInRotation})  is not very
small. The rotational contribution to the normal component density
normalized to the  density of the $^3$He atoms is $\Delta n_{{\rm
n}\parallel}/n =3\Omega/m_3 c_\perp^2$ which is $ \sim 10^{-4}$ for
$\Omega \sim 3$ rad/s. This is within the
resolution of the vibrating wire detectors.

We finally mention a possible application of our results to the
 superconducting Sr$_2$RuO$_4$ \cite{Rice}.
An advantage of using superconductors is that the particle current
$\Delta{\bf J}_{\rm q}$ in Eq.(\ref{Current})  is accompanied by the
electric current $e\Delta{\bf J}_{\rm q}$, and can be measured directly.
An observation in Sr$_2$RuO$_4$ of the analogue of the parity violating
effect that we discussed here  (or of the other effects coming from the
induced Chern-Simons terms \cite{Goryo}), would be an unquestionable
evidence of the chirality of this superconductor.

We thank Matti Krusius who suggested to apply our results to chiral
superconductors.  The work of G.V. was supported in part by the Russian
Foundation for Fundamental Research grant No. 96-02-16072 and by European
Science Foundation. The work of A.V. was supported in part by the National
Science Foundation.

\end{document}